\documentclass[twoside,12pt]{article}
\newcommand\SEKImasterusepackages{}
\newcommand\SEKIusersusepackages
{
}

\newcommand\SEKIREVIEWSTATUS{1}

\newcommand\SEKISUBEDITOR
{Erica Melis 
\\FR Informatik, Universit\"at des Saarlandes, D--66123 Saarbr\"ucken, Germany
\\E-mail: {\tt melis@activemath.org}
\\WWW: \url{http://www.activemath.org/~melis/}
}

\newcommand\SEKIREVIEWER
{Claus-Peter Wirth
\\E-mail: \url{cp@ags.uni-sb.de}
\\WWW: \url{http://www.ags.uni-sb.de/~cp/welcome.html}
}

\newcommand\SEKIREFEREEPROCESS
{
}

\newcommand\SEKIYEAR{99}

\newcommand\SEKINUMBEROFYEAR{02}

\newcommand\SEKITYPE{0}

\newcommand\SEKIPUBLISHEDTYPE{0}
\newcommand\SEKIPUBLISHEDAS
{}
\date{1999}

\title{Resource Adaptive Agents in Interactive Theorem Proving}

\author
{
Christoph Benzm\"uller and Volker Sorge \\ 
FR Informatik, Saarland Univiversity, Germany \url{{chris|sorge}@ags.uni-sb.de}\\
}

\input seki-deckblatt

\usepackage{mathptmx}       
\usepackage{helvet}         
\usepackage{courier}        
\usepackage{type1cm}        
\usepackage{makeidx}         
\usepackage{graphicx}        
\usepackage{multicol}        
\usepackage[bottom]{footmisc}

\usepackage{latexsym}
\usepackage{longtable}
\usepackage{wrapfig}
\usepackage{verbatim}
\usepackage{url}
\usepackage{xspace}

\usepackage{amssymb}  
\usepackage{proof}

\newenvironment{ednote}{\begin{svgraybox}\noindent\textbf{EdNote:}\par}
  {\end{svgraybox}}

\newcommand{\LOUI}{\textsc{L$\Upomega$ui}\xspace}

\newcommand{\OMEGA}{\textsc{$\Upomega$mega}\xspace}

\newcommand{\database}[1]{
\rput(0,1.5){\rnode{h1}{\psellipse(1,0)(1,0.3)}}  
\rput(0,0){\psline{-}(0,1.5)}
\rput(2,0){\psline{-}(0,1.5)}
\rput(0,0){\rnode{h2}{\psline[linearc=0.7]{-}(0,0)(0.2,-0.12)(0.6,-0.23)(1,-0.26)(1.4,-0.23)(1.8,-0.12)(2,0)}}
}

\def\ambnormform#1{{#1}\hspace*{-1.1ex}\downarrow_{\kern-.2em\scriptscriptstyle *}} 

\def\tt{STT}

\def\omega{{$\Omega${\sc mega}}}
\def\OMEGA{{$\Omega${\sc mega}}}
\def\loui{{${\cal L}\Omega{\cal UI}$}}
\def\LOUI{\loui}
\def\hol{{\sc Hol}}

\def\tps{{\sc Tps}}

\def\agent{{\mathfrak A}}  
\def\cagent{{\mathfrak C}} 
\def\pline#1#2{{{#1}_{#2}}}
\def\cT{{\cal T}}

\newif\ifgerman\germanfalse
\newif\iffrench\frenchfalse
\newcommand{\JAR}{Journal of Automated Reasoning}
\newcommand{\PROC}{{Proceedings}}
\newcommand{\LNAI}{LNAI}
\newcommand{\LNCS}{LNCS}
\def\NETHERLANDS{\ifgerman{Niederlande}\else{\iffrench{Pays Bas}\else{The Netherlands}\fi}\fi}
\def\JULY{\ifgerman{Juli }\else{\iffrench{juillet }\else{July }\fi}\fi}
\def\NOVEMBER{\ifgerman{November }\else{\iffrench{novembre }\else{November }\fi}\fi}

\def\deq{{\;\colon\kern-.1em=\;}}
\makeatletter\renewcommand{\ednote}[2][]{\ed@note{#2}{E}{#1}}\makeatother

\begin{document}
\makecover


\begin{abstract}
  We introduce a resource adaptive agent mechanism which supports the user in interactive theorem
  proving. The mechanism, an extension of~\cite{BeSo98b}, uses a two layered architecture of agent
  societies to suggest appropriate commands together with possible command argument instantiations.
  Experiments with this approach show that its effectiveness can be further improved by introducing
  a resource concept.  In this paper we provide an abstract view on the overall mechanism, motivate
  the necessity of an appropriate resource concept and discuss its realization within the agent
  architecture.

\end{abstract}

\section{Introduction}
\label{sec:intro}

Interactive theorem provers have been developed to overcome the shortcomings of purely
automatic systems by enabling the user to guide the proof search and by directly importing
expert knowledge into the system. For large proofs, however, this task becomes difficult
if the system does not provide a sophisticated mechanism to suggest possible next steps in
order to minimize the necessary interactions.

Suggestion mechanisms in interactive theorem proving systems such as
\hol~\cite{GoMe93}, \tps~\cite{Andrews96} or \omega~\cite{Omega97}
are rather limited in their functionality as they usually

\hspace{-.5cm}\begin{tabular}{lp{11cm}}
(i)   & use inflexible sequential computation strategies,\\
(ii)  & do not have anytime character,\\
(iii) & do not work steadily and autonomously in the background of a system, and\\ 
(iv)  & do not exhaustively use available computation resources
\end{tabular}

In order to overcome these limitations, \cite{BeSo98b} proposed --- within the
con\-text\footnote{We want to point out that this mechanism is in no way restricted to a
  specific logic or calculus and can easily be adapted to other interactive theorem
  contexts as well.} of tactical theorem proving based on ND-cal\-cu\-lus~\cite{Gentzen35}
--- a new, flexible support mechanism with anytime character. It suggests commands,
applicable in the current proof state --- more precisely commands that invoke some ND-rule
or tactic --- together with a suitable argument instantiations.  It is based on two layers
of societies of autonomous, concurrent agents which steadily work in the background of a
system and dynamically update their computational behavior to the state of the proof
and/or specific user queries to the suggestion mechanism. By exchanging relevant results
via blackboards the agents cooperatively accumulate useful command suggestions which can
then be heuristically sorted and presented to the user.

A first Lisp-based implementation of the support mechanism in the \omega-system yielded promising
results.  However as we could not exploit concurrency in the Lisp implementation experience showed
that we had to develop a resource adapted concept for the agents in order to allow for efficient
suggestions even in large examples. In the current reimplementation of major parts of {\omega} and
the command suggestion mechanism in Oz, a concurrent constraint logic programming
language~\cite{Smolka:TOPM95}, the first impression is that the use of concurrency provides a good
opportunity to switch from a static to a dynamic, resource adaptive control of the mechanism's
computational behavior. Thereby, we can exploit both knowledge on the prior performance of the
mechanism as well as knowledge on classifying the current proof state and single agents in order to
distribute resources.

In this paper we first sketch our two layered agent mechanism and introduce the static approach of
handling resources. We then extend this approach into a resource adaptive\footnote{In this paper we
  adopt the notions of resource adapted and resource adaptive as defined in~\cite{Zilberstein95},
  where the former means that agents behave with respect to some initially set resource
  distribution. According to the latter concept agents have an explicit notion of resources
  themselves, enabling them to actively participate in the dynamic allocation of resources.} one
where the agents monitor their own contributions and performance in the past in order to estimate
their potential contribution and performance within the next step and where the adequacy of all
agents computations with respect to the complexity of the current proof goal is analyzed by a
special {\it classification agent}.  Thus, the agents have a means to decide whether or not they
should pursue their own intentions in a given state.  Moreover, the overall mechanism can
dynamically change resource allocations thereby enabling or disabling computations of single agents
or complete societies of agents.  And finally, the classification agent sends deactivation signals
to single agents or agent societies as soon as evidence or definite knowledge is available that these
agents can currently not compute any appropriate contributions (e.g.~when an agent belongs to a
first-order command whereas the current proof goal can be classified as propositional).

\section{Reference Example}
The following example will be used throughout this paper: $(p_{o\rightarrow o}\ (a_o \wedge b_o)) \Rightarrow
(p\ (b \wedge a))$. Whereas it looks quite simple at a first glance this little higher-order (HO) problem
cannot be solved by most automatic HO theorem provers known to the authors, since it requires the application
of the extensionality principles which are generally not built-in in HO theorem proving systems.
Informally this example states: If the
truth value of $a \wedge b$ is element of the set $p$ of truth values, then $b \wedge a$ is also in $p$.  In
the mathematical assistant {\omega}~\cite{Omega97}, which employs a variant of Gentzen's natural deduction
calculus (ND)~\cite{Gentzen35} enriched by more powerful proof tactics, the following proof can easily be
constructed interactively\footnote{Linearized ND proofs are presented as described in~\cite{Andrews86}.  Each
  proof line consists of a label, a set of hypotheses, the formula and a justification.}:\\[.2cm]
\begin{small}
$\begin{array}{ll@{\;\vdash\;\;}ll} 
L_1 & \hspace{-.1cm} (L_1) & (p\ (a \wedge b))   &   \mbox{Hyp} \\ 
L_4 & \hspace{-.1cm} (L_1) & (b \wedge a) \Leftrightarrow (a \wedge b)   & \mbox{\sc Otter}\\ 
L_3 & \hspace{-.1cm} (L_1) & (b \wedge a) = (a \wedge b)   &  \mbox{$\Leftrightarrow$2$=$}:
(L_4) \\ 
L_2 & \hspace{-.1cm} (L_1) & (p\ (b \wedge a))  &  =_{\mbox{\tiny subst}}: (\langle 1\rangle)
(L_1 L_3)\\ 
C & \hspace{-.1cm} () & (p\ (a \wedge b)) \Rightarrow (p\ (b \wedge a)) & \Rightarrow_I: (L_2)\\
\end{array}$
\end{small}\\[.2cm]
The idea of the proof is to show that the truth value of $a \wedge b$ equals that of $b \wedge a$
(lines $L_3$ and $L_4$) and then to employ equality substitution (line $L_2$). The equation $(b
\wedge a) = (a \wedge b)$, i.e.\ by boolean extensionality the equivalence $(b \wedge a)
\Leftrightarrow (a \wedge b)$ is proven here with the first-order prover {\sc
  Otter}~\cite{McCWos:occi97} and the detailed subproof is hidden at the chosen presentation level.
Our agent mechanism is able to suggest all the single proof steps together with the particular
parameter instantiations to the user. In this paper we use parts of this example to demonstrate the
working scheme of the mechanism and to motivate the use of resources.

\section{Suggesting Commands}

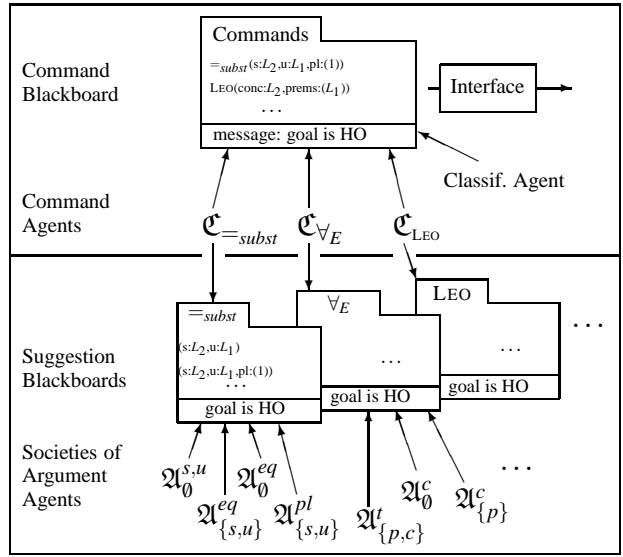
\begin{figure}
\begin{center}
\setlength{\unitlength}{3947sp}%
\begingroup\makeatletter\ifx\SetFigFont\undefined
\def\x#1#2#3#4#5#6#7\relax{\def\x{#1#2#3#4#5#6}}%
\x\fmtname xxxxxx\relax \def\y{splain}%
\ifx\x\y   
\gdef\SetFigFont#1#2#3{%
  \ifnum #1<17\tiny\else \ifnum #1<20\small\else
  \ifnum #1<24\normalsize\else \ifnum #1<29\large\else
  \ifnum #1<34\Large\else \ifnum #1<41\LARGE\else
     \huge\fi\fi\fi\fi\fi\fi
  \csname #3\endcsname}%
\else
\gdef\SetFigFont#1#2#3{\begingroup
  \count@#1\relax \ifnum 25<\count@\count@25\fi
  \def\x{\endgroup\@setsize\SetFigFont{#2pt}}%
  \expandafter\x
    \csname \romannumeral\the\count@ pt\expandafter\endcsname
    \csname @\romannumeral\the\count@ pt\endcsname
  \csname #3\endcsname}%
\fi
\fi\endgroup
\begin{picture}(3849,3474)(64,-2698)
  \thinlines \put( 76,-2686){\framebox(3825,3450){}} \put( 76,-811){\line( 1, 0){1200}}
  \put(1426,-811){\line( 1, 0){450}} \put(2026,-811){\line( 1, 0){450}} \put(2626,-811){\line( 1,
    0){1275}} \put(1276,-136){\line( 0, 1){825}} \put(1276,689){\line( 1, 0){600}}
  \put(1876,689){\line( 1, 0){150}} \put(2026,689){\line( 0,-1){150}} \put(2026,539){\line( 1,
    0){600}} \put(2626,539){\line( 0,-1){675}} \put(2626,-136){\line(-1, 0){1350}}
  \put(2701,239){\line( 1, 0){75}} \put(2776,239){\line( 0, 1){150}} \put(2776,389){\line( 1,
    0){600}} \put(3376,389){\line( 0,-1){300}} \put(3376, 89){\line(-1, 0){600}} \put(2776,
  89){\line( 0, 1){150}} \put(3376,239){\vector( 1, 0){225}} \put(1126,-1261){\line( 0,-1){600}}
  \put(1126,-1861){\line( 1, 0){900}} \put(2026,-1861){\line( 0, 1){600}} \put(2026,-1261){\line(-1,
    0){450}} \put(1576,-1261){\line( 0, 1){150}} \put(1576,-1111){\line(-1, 0){450}}
  \put(1126,-1111){\line( 0,-1){150}} \put(1876,-1261){\line( 0, 1){225}} \put(1876,-1036){\line( 1,
    0){525}} \put(2401,-1036){\line( 0,-1){150}} \put(2401,-1186){\line( 1, 0){375}}
  \put(2776,-1186){\line( 0,-1){600}} \put(2776,-1786){\line(-1, 0){750}} \put(2626,-1186){\line( 0,
    1){225}} \put(2626,-961){\line( 1, 0){450}} \put(3076,-961){\line( 0,-1){150}}
  \put(3076,-1111){\line( 1, 0){450}} \put(3526,-1111){\line( 0,-1){600}} \put(3526,-1711){\line(-1,
    0){750}} \put(2926,-2236){\vector(-1, 2){225}} \put(2626,-2161){\vector(-1, 3){127.500}}
  \put(2326,-2386){\vector( 0, 1){600}} \put(1651,-2086){\vector(-1, 3){75}}
  \put(1426,-2311){\vector( 0, 1){450}} \put(1201,-2086){\vector( 1, 3){75}}
  \put(1351,-736){\vector( 0,-1){375}} \put(1351,-511){\vector( 1, 4){92.647}}
  \put(1951,-511){\vector( 0, 1){375}} \put(1951,-736){\vector( 0,-1){300}} \put(2551,-736){\vector(
    1,-3){75}} \put(2551,-511){\vector(-1, 4){92.647}} \put(1876,-2311){\vector(-1, 4){110.294}}
  \put(1126,-1711){\line( 1, 0){900}} \put(2026,-1636){\line( 1, 0){750}} \put(2776,-1561){\line( 1,
    0){750}} \put(1276, 14){\line( 1, 0){1350}} \put(3016,-219){\vector(-2, 1){375.200}}
  \put(3601,-1261){\makebox(0,0)[lb]{\smash{\SetFigFont{12}{16}{rm}$\ldots$}}}
  \put(3151,-1411){\makebox(0,0)[lb]{\smash{\SetFigFont{8}{9.3}{rm}$\ldots$}}}
  \put(2401,-1486){\makebox(0,0)[lb]{\smash{\SetFigFont{8}{9.3}{rm}$\ldots$}}}
  \put(3151,-2161){\makebox(0,0)[lb]{\smash{\SetFigFont{12}{16}{rm}$\ldots$}}}
  \put(151,-2086){\makebox(0,0)[lb]{\smash{\SetFigFont{8}{9.3}{rm}Societies of}}}
  \put(151,-2236){\makebox(0,0)[lb]{\smash{\SetFigFont{8}{9.3}{rm}Argument}}}
  \put(151,-2386){\makebox(0,0)[lb]{\smash{\SetFigFont{8}{9.3}{rm}Agents}}}
  \put(151,-1486){\makebox(0,0)[lb]{\smash{\SetFigFont{8}{9.3}{rm}Suggestion}}}
  \put(151,-1636){\makebox(0,0)[lb]{\smash{\SetFigFont{8}{9.3}{rm}Blackboards}}}
  \put(1351,-1200){\makebox(0,0)[b]{\smash{\SetFigFont{8}{9.3}{rm}$=_{subst}$}}}
  \put(2851,-1090){\makebox(0,0)[b]{\smash{\SetFigFont{8}{9.3}{rm}{\sc Leo}}}}
  \put(3076,209){\makebox(0,0)[b]{\smash{\SetFigFont{8}{9.3}{rm}Interface}}}
  \put(1651,539){\makebox(0,0)[b]{\smash{\SetFigFont{8}{9.3}{rm}Commands}}}
  \put(151,314){\makebox(0,0)[lb]{\smash{\SetFigFont{8}{9.3}{rm}Command}}}
  \put(151,-511){\makebox(0,0)[lb]{\smash{\SetFigFont{8}{9.3}{rm}Command}}}
  \put(151,-661){\makebox(0,0)[lb]{\smash{\SetFigFont{8}{9.3}{rm}Agents}}}
  \put(151,164){\makebox(0,0)[lb]{\smash{\SetFigFont{8}{9.3}{rm}Blackboard}}}
  \put(2146,-1150){\makebox(0,0)[b]{\smash{\SetFigFont{8}{9.3}{rm}$\forall_E$}}}
  \put(1026,-2270){\makebox(0,0)[lb]{\smash{\SetFigFont{11}{14}{rm}$\agent^{s,u}_{\emptyset}$}}}
  \put(1250,-2500){\makebox(0,0)[lb]{\smash{\SetFigFont{11}{14}{rm}$\agent^{eq}_{\{s,u\}}$}}}
  \put(1500,-2270){\makebox(0,0)[lb]{\smash{\SetFigFont{11}{14}{rm}$\agent^{eq}_{\emptyset}$}}}
  \put(1730,-2500){\makebox(0,0)[lb]{\smash{\SetFigFont{11}{14}{rm}$\agent^{pl}_{\{s,u\}}$}}}
  \put(2250,-2530){\makebox(0,0)[lb]{\smash{\SetFigFont{11}{14}{rm}$\agent^{t}_{\{p,c\}}$}}}
  \put(2530,-2350){\makebox(0,0)[lb]{\smash{\SetFigFont{11}{14}{rm}$\agent^{c}_{\emptyset}$}}}
  \put(2870,-2400){\makebox(0,0)[lb]{\smash{\SetFigFont{11}{14}{rm}$\agent^{c}_{\{p\}}$}}}
  \put(2476,-680){\makebox(0,0)[lb]{\smash{\SetFigFont{12}{16}{rm}$\cagent_{\mbox{{\tiny {\sc
                Leo}}}}$}}}
  \put(1876,-680){\makebox(0,0)[lb]{\smash{\SetFigFont{12}{16}{rm}$\cagent_{\forall_E}$}}}
  \put(1276,-680){\makebox(0,0)[lb]{\smash{\SetFigFont{12}{16}{rm}$\cagent_{=_{subst}}$}}}
\put(1651, 89){\makebox(0,0)[lb]{\smash{\SetFigFont{8}{9.3}{rm}$\ldots$}}}
\put(1330,220){\makebox(0,0)[lb]{\smash{\SetFigFont{5}{5}{rm}{\sc Leo}(conc:$L_2$,prems:($L_1$))}}}
\put(1330,370){\makebox(0,0)[lb]{\smash{\SetFigFont{5}{5}{rm}$=_{subst}$(s:$L_2$,u:$L_1$,pl:(1))}}}
\put(1130,-1411){\makebox(0,0)[lb]{\smash{\SetFigFont{5}{5}{rm}(s:$L_2$,u:$L_1$)}}}
\put(1130,-1561){\makebox(0,0)[lb]{\smash{\SetFigFont{5}{5}{rm}(s:$L_2$,u:$L_1$,pl:(1))}}}
\put(1426,-1636){\makebox(0,0)[lb]{\smash{\SetFigFont{8}{9.3}{rm}$\ldots$}}}
\put(1300,-1809){\makebox(0,0)[lb]{\smash{\SetFigFont{7}{8.4}{rm}goal is HO}}}
\put(2086,-1733){\makebox(0,0)[lb]{\smash{\SetFigFont{7}{8.4}{rm}goal is HO}}}
\put(2829,-1666){\makebox(0,0)[lb]{\smash{\SetFigFont{7}{8.4}{rm}goal is HO}}}
\put(1350,-84){\makebox(0,0)[lb]{\smash{\SetFigFont{7}{8.4}{rm}message: goal is HO}}}
\put(2799,-383){\makebox(0,0)[lb]{\smash{\SetFigFont{8}{9.3}{rm}Classif.~Agent}}}
\end{picture}
\end{center}
\caption{The two layered suggestion mechanism\label{fig:suggestions}.}\vspace{-.5cm}
\end{figure}

The general suggestion mechanism is based on a two layered agent architecture displayed in
Fig.~\ref{fig:suggestions} which shows the actual situation after the first proof step in the
example, where the backward application of $\Rightarrow_I$ introduces the hypothesis line $L_1$ and
$L_2$ as the new open goal. The task of the bottom layer of agents (cf.~the lower part of
Fig.~\ref{fig:suggestions}) is to compute possible argument instantiations for the provers commands
in dependence of the dynamically changing partial proof tree. The task of the top layer (cf.~the
upper part of Fig.~\ref{fig:suggestions}) is to collect the most appropriate suggestions from the
bottom layer, to heuristically sort them and to present them to the user.
  
The bottom layer consists of societies of argument agents where each society belongs to
exactly one command associated with a proof tactic\footnote{For our approach it is not
  necessary to distinguish between proof rules, proof tactics and proof methods and
  therefore we just use the phrase "tactic" within this article.} (cf.\ below).  On the
one hand each argument agent has its own intention, namely to search in the partial proof
for a proof line that suits a certain specification. On the other hand argument agents
belonging to the same society also pursue a common goal, e.g.\ to cooperatively compute
most complete argument suggestions (cf.\ concept of PAI's below) for their associated
command.  Therefore the single agents of a society exchange their particular results via a
suggestion blackboard and try to complete each others suggestions.

The top layer consists of a single society of command agents which steadily 
monitor the particular suggestion blackboards on the bottom layer.
For each suggestion blackboard there exists one command agent whose intention is to
determine the most complete suggestions and to put them on the command blackboard.

The whole distributed agent mechanism runs always in the background of the interactive
theorem proving environment thereby constantly producing command suggestions that are
dynamically adjusted to the current proof state.  At any time the suggestions on the
command blackboard are monitored by an interface component which presents them
heuristically sorted to the user via a graphical user interface.  As soon as the user
executes a command the partial proof is updated and simultaneously the suggestion and
command blackboards are reinitialized.



\subsection{Partial Argument Instantiations (PAI)}

The data that is exchanged within the blackboard architecture heavily depends 
on a concept called a  {\em partial argument instantiation (PAI)}\/ of a
command. In order to clarify our mechanism we need to introduce this concept
in detail.  

In an interactive theorem prover such as \OMEGA\/ or \hol\/ one has generally one command
associated with each proof tactic that invokes the application of this tactic to a set of
proof lines.  In \OMEGA\/ these tactics have a fixed {\em outline}, i.e.\ a set of premise
lines, conclusion lines and additional parameters, such as terms or term-positions. Thus
the general instance of a tactic $\cT$ can be formalized in the following way:
\[\infer[{\cal T} (Q_1\cdots Q_n)]{C_1\cdots C_m}{P_1\cdots P_l},\]
where we call the $P_i, C_j, Q_k$ the {\em formal arguments}\/ of the tactic $\cT$
(we give an example below).

We can now denote the command {\it t}\/ invoking tactic $\cT$ formally in a similar fashion as
\[\infer[{\it t} (q_{k_1}\cdots q_{k_{n^\prime}})]{c_{j_1}\cdots
  c_{j_{m^\prime}}}{p_{i_1}\cdots p_{i_{l^\prime}}},\] where the formal arguments $p_i, c_j, q_k$ of {\em t}\/
correspond to a subset of the formal arguments of the tactic. 
To successfully execute the command some, not necessarily all, formal arguments have to be instantiated with
{\em actual arguments}, e.g.\ proof lines. A set of pairs relating each formal argument of the command to an
(possibly empty) actual argument is called a {\em partial argument instantiation (PAI)}.

We illustrate the idea of a PAI using the tactic for {\em equality substitution} ${=}_{Subst}$ and its
corresponding command {\it {=}Subst}\/ as an example.


\[
\infer[{=}_{Subst} (P^*)]{\Phi^\prime[y]}{\Phi[x] & x=y}
\qquad \longrightarrow \qquad
\infer[{{=}Subst} (pl)]{s}{u & eq}
\]

\noindent Here $\Phi[s]$ is an arbitrary higher order formula with at least one
occurrence of the term $s$, $P^*$ is a list of term-positions representing one or several
occurrences of $s$ in $\Phi$, and $\Phi^\prime[t]$ represents the term resulting from replacing $s$
by $t$ at all positions $P^*$ in $\Phi$.  {\it u, eq, s}\/ and {\it pl}\/ are the corresponding
formal arguments of the command associated with the respective formal arguments of the tactic. We
observe the application of this tactic to line $L_2$ of our example:

\begin{small}
\[\begin{array}{lllll} 
L_1 & (L_1) & \vdash & (p\ (a \wedge b))   &   \mbox{Hyp} \\ 
    &       &    $\dots$ &                     & \\
L_2 & (L_1) & \vdash & (p\ (b \wedge a))  &  \mbox{Open} \\ 
\end{array}\]
\end{small}

\noindent
As one possible PAI for {\it {=}Subst}\/ we get the set of pairs $(u{:}\pline{L}{1}, eq{:}\epsilon,
s{:}\pline{L}{2}, pl{:}\epsilon)$, where $\epsilon$ denotes the empty or unspecified actual argument. We omit
writing pairs containing $\epsilon$ and, for instance, write the second possible PAI of the above example as
$(u{:}\pline{L}{1}, s{:}\pline{L}{2}, pl{:}(\langle 1\rangle))$. To execute {\it =Subst}\/ with the former PAI
the user would have to at least provide the position list, whereas using the latter PAI results in the line
$\pline{L}{3}$ of the example containing the equality.

\subsection{Argument Agents}
The idea underlying our mechanism to suggest commands is to compute PAIs as complete as possible for each
command, thereby gaining knowledge on which tactics can be applied combined
with which argument instantiations in a given proof state. 

The main work is done by the societies of cooperating {\em Argument Agents}\/ at the bottom layer (cf.\ 
Fig.~\ref{fig:suggestions}).  Their job is to retrieve information from the current proof state either by
searching for proof lines appropriate to the agents specification or by computing some additional parameter
(e.g.\ a list of sub-term positions) with already given information. Sticking to our example we can informally
specify the agents $\agent^{u,s}_{\emptyset}$, $\agent^{eq}_{\emptyset}$, $\agent^{eq}_{\{u,s\}}$, and
$\agent^{pl}_{\{u,s\}}$ for the {\it =Subst}\/ command (cf.~\cite{BeSo98b} for a formal specification):
\begin{eqnarray*}
  \agent^{s,u}_{\emptyset} \hspace{.25cm} & \hspace{-.4cm}{=} &\hspace{-.4cm}
  \left\{\parbox{6.7cm}{\small find open line and a support line that differ only wrt.\ occurences of a single proper subterm}\right\}\\
  \agent^{eq}_{\emptyset}  \hspace{.35cm} & \hspace{-.4cm}{=} &\hspace{-.4cm} \{\mbox{\small find line with an equality}\}\\
  \agent^{eq}_{\{u,s\}} & \hspace{-.4cm}{=} &\hspace{-.4cm} \left\{\mbox{\small find equality line suitable for {\it s}\/ and {\it u}}\right\}\\
  \agent^{pl}_{\{u,s\}} & \hspace{-.4cm}{=} &\hspace{-.4cm} \left\{\mbox{\small compute positions where {\it s}\/ and {\it u}\/ differ}\right\}
\end{eqnarray*}
The attached superscripts specify the formal arguments of the command for which actual arguments are computed,
whereas the indices denote sets of formal arguments that necessarily have to be instantiated in some PAI, so
that the agent can carry out its own computations. For example agent $\agent^{eq}_{\{u,s\}}$ only starts
working when it detects a PAI on the blackboard where actual arguments for {\it u}\/ and {\it s}\/ have been
instantiated. On the contrary $\agent^{eq}_{\emptyset}$ does not need any additional knowledge in order to
pursue its task to retrieve an open line containing an equality as formula.

The agents themselves are realized as autonomous processes that concurrently compute their
suggestions whereas they are triggered by the PAIs on the blackboard, i.e.\ the results of
other agents of their society. For
instance both agents $\agent^{eq}_{\{u,s\}}$ and $\agent^{pl}_{\{u,s\}}$ would simultaneously 
start their search as
soon as $\agent^{s,u}_{\emptyset}$ has returned a result. The agents of one society 
cooperate in the sense that they activate each other (by writing new PAIs to the 
blackboard) and furthermore complete each others suggestions.

Conflicts between agents do not arise, as agents that add actual parameters to some PAI always write
a new copy of the particular PAI on the blackboard, thereby keeping the original less complete PAI
intact.  The agents themselves watch their suggestion blackboard (both PAI entries and additional
messages, cf.~\ref{sec:resources}) and running agents terminate as soon as the associated suggestion
blackboard is reinitialized, e.g.\ when a command has been executed by the user.

The left hand side of Fig.~\ref{fig:suggestions} illustrates our above example:
The topmost suggestion blackboard contains the two PAIs:
$(u{:}\pline{L}{1}, s{:}\pline{L}{2})$ computed by agent $\agent^{s,u}_{\emptyset}$ and
$(u{:}\pline{L}{1}, s{:}\pline{L}{2}, pl{:}(\langle 1\rangle))$ completed by agent $\agent^{eq}_{\{u,s\}}$.


\subsection{Command Agents}
In the society of {\em command agents}\/ every agent is linked to a command and its task
is to initialize and monitor the associated suggestion blackboards. Its intention is to
select among the entries of the associated blackboard the most complete and appropriate one and
to pass it, enriched by the corresponding command name to the command blackboard.  That
is, as soon as a PAI is written to the related blackboard that has at least one actual
argument instantiated, the command agent suggests the command as applicable in the current
proof state, providing also the PAI as possible argument instantiations. It then updates
this suggestion, whenever a {\em better}\/ PAI has been computed. In this context {\em
  better}\/ generally means a PAI containing more actual arguments. In the case of our
example the current PAI suggested by command agent $\cagent_{=Subst}$ is
$(u{:}\pline{L}{1}, s{:}\pline{L}{2}, pl{:}(\langle 1\rangle))$.

These suggestions are accumulated on a {\em command blackboard}, that simply stores all suggested
commands together with the proposed PAI, continuously handles updates of the latter, sorts and resorts the single
suggestions and provides a means to propose them to the user. In the case of the \OMEGA-system this is achieved
in a special command suggestion window within the graphical user interface \LOUI~\cite{LOUI98}. The sorting of
the suggestions is done according to several heuristic criteria, one of which is that commands
with fully instantiated PAIs are always preferred as their application may conclude the whole subproof.


\subsection{Experiences}
The mechanism sketched here has some obvious advantages.  Firstly, it has anytime
character and does not waste system resources, as it constantly works in a background
process and steadily improves its suggestions wrt.\ its heuristic sorting
criteria. Secondly, in contrast to conventional command and argument suggestion
mechanisms, our approach provides more flexibility in the sense that agents can be defined
independently from the actual command and the user can choose freely among several given
suggestions. Thirdly, we can employ concurrent programming techniques to parallelize the
process of computing suggestions.

Unfortunately, computations of single agents themselves can be very costly. Reconsider the agents of command
{\it =Subst}\/: In \OMEGA\/ we have currently 25 different argument agents defined for {\it =Subst}\/ where
some are computationally highly expensive. For example, while the agent $\agent^{eq}_{\emptyset}$ only tests
head symbols of formulas during its search for lines containing an equality and is therefore relatively
inexpensive, the agent $\agent^{u,s}_{\emptyset}$ performs computationally expensive matching operations.  In
large proofs agents of the latter type might not only take a long time before returning any useful result, but
also will absorb a fair amount of system resources, thereby slowing down the computations of other argument
agents.

\cite{BeSo98b} already tackles this problem partially by introducing a {\em focusing
  technique}\/ that explicitly partitions a partial proof into subproblems in order to
guide the search of the agents. This focusing technique takes two important aspects into
account: (i) A partial proof often contains several open subgoals and humans usually focus
on one such subgoal before switching to the next. (ii) Hypotheses and derived lines
belonging to an open subgoal are chronologically sorted where the interest focuses on the
more recently introduced lines. Hence, the agents restrict their search to the actual
subgoal ({\em actual focus}\/) and guide their search according to the chronological order
of the proof lines.

\section{Resource Adapted Approach}\label{sec:resources}
Apart from this we use a concept of {\em static complexity ratings}\/ where a rating is attached to
each argument and each command agent, that roughly reflects the computational complexity involved
for its suggestions. A {\em global complexity value}\/ can then be adjusted by the user permitting
to suppress computations of agents, whose ratings are larger than the specified value. Furthermore
commands can be completely excluded from the suggestion process. For example, the agent
$\agent^{u,s}_{\emptyset}$ has a higher complexity rating than $\agent^{eq}_{\emptyset}$ from the
{\it =Subst}\/ example, since recursively matching terms is generally a harder task than retrieving
a line containing an equality. The overall rating of a command agent is set to the average rating of
its single argument agents. This rating system increased the effectiveness of the command
suggestions for the simulation of the architecture in LISP.  (In the simulation a user was always
forced to wait for all active agents to finish their computations.)  However, the system is very
inflexible as ratings are assigned by the programmer of a particular agent and cannot be adjusted by
the user at runtime.  Furthermore, the ratings do not necessarily reflect the actual relative
complexity of the agents.

Shifting to the Oz implementation the latter problem no longer arises since a user can interrupt the
suggestion process by choosing a command at any time without waiting for all possible suggestions to be made.
In the Oz implementation every agent is an independent thread. It either quits its computations regularly or
as soon as it detects that the blackboard it works for has been reinitialized, i.e.\ when the user has
executed a command. It then performs all further computations with respect to the reinitialized blackboard.

However, with increasing size of proofs some agents never have the chance to write meaningful suggestions to a
blackboard. Therefore, these agents should be excluded from the suggestion process altogether, especially if
their computations are very costly which deprives other agents of resources. But while in the sequential
simulation in LISP it was still possible for a user to monitor which agent is computationally very expensive
and to subsequently adjust the global complexity value accordingly, this is no longer possible in the
concurrent setting as the user in general is not aware of which agents produced useful results. Furthermore,
the user should be as far as possible spared from fine-tuning a mechanism designed for his/her own support.

\section{Resource Adaptive Approach}
In this section we extend the resource adapted approach into a resource adaptive one.  While
we retain the principle of activation/deactivation by comparing the particular complexity ratings of the
argument agents with the overall deactivation threshold, we now allow the individual complexity ratings of
argument agents to be dynamically adjusted by the system itself.
Furthermore, we introduce a special classification agent which analyses and classifies 
the current proof goal in order to deactivate those agents which are not appropriate 
wrt.~the current goal. 

\subsection{Dynamic Adjustment of Ratings}
The dynamic adjustment takes place on both layers: On the bottom layer we allow the argument agents to adjust
their own ratings by reflecting their performance and contributions in the past.  On the other hand the
command agents on the top layer adjust the ratings of their associated argument agents. This is motivated by
the fact that on this layer it is possible to compare the performance and contribution of agent societies
of the bottom layer. 

Therefore, agents need an explicit concept of resources enabling them to communicate and reason about their
performance. The communication is a\-chieved by propagating resource informations from the bottom to the top
layer and vice versa via the blackboards. The actual information is gathered by the agents on the bottom layer
of the architecture. Currently the argument agents evaluate their effectiveness with respect to the following
two measures:

\begin{enumerate}
\item \label{time} the absolute cpu time the agents consume, and
\item \label{patience}\lq the patience of the user\rq,  before executing the next command.
\end{enumerate}
(\ref{time}) is an objective measure that is computed by each agent at runtime. Agents then 
use these values to compute the average cpu time for the last $n$ runs and convert the result into a
corresponding complexity rating.





Measure (\ref{patience}) is rather subjective which expresses formally the ability of an agent to judge
whether it ever makes contributions for the command suggesting process in the current proof state.  Whenever
an agent returns from a computation without any new contribution to the suggestion blackboard, or even worse,
whenever an agent does not return before the user executes another command (which reinitializes the
blackboards), the agent receives a penalty that increases its complexity rating.  Consequently, when an agent
fails to contribute several times in a row, its complexity rating quickly exceeds the deactivation
threshold and the agent retires.

Whenever an argument agent updates its complexity rating this adjustment is reported to the corresponding
command agent via a blackboard entry. The command agent collects all these entries, computes the average
complexity rating of his argument agents, and reports the complete resource information on his society of
argument agents to the command blackboard. The command blackboard therefore steadily provides
information on the effectiveness of all active argument agents, as well as information on 
the retired agents and an estimation of the overall effectiveness every argument agent society.

An additional {\em resource agent}\/ uses this resource information in order to reason about a possibly
optimal resource adjustment for the overall system, taking the following criteria into account:
\begin{itemize}
\item Assessment of absolute cpu times.
\item A minimum number of argument agents should always be active. If the number of active agents drops
  below this value the global complexity value is readjusted in order to reactivate some of the retired
  agents.
\item Agent societies with a very high average complexity rating and many retired argument agents should get a
  new chance to improve their effectiveness. Therefore the complexity ratings of the retired agents is lowered
  beneath the deactivation threshold.
\item In special proof states some command agent (together with their argument agents) are excluded. For
  example, if a focused subproblem is a propositional logic problem, commands invoking tactics dealing with
  quantifiers are needless.
\end{itemize}

Results from the resource agent are propagated down in the agent society and gain precedence over
the local resource adjustments of the single agents.

\subsection{Informed Activation \& Deactivation}
Most tactics in an interactive theorem prover are implicitly associated with a specific logic
(e.g.~propositional, first-order, or higher -order logic) or even with a specific mathematical
theory (e.g.~natural numbers, set theory). This obviously also holds for the proof problems examined
in a mathematical context. Some systems -- for instance the {\omega}-System -- do even explicitly
maintain respective knowledge by administering all rules, tactics, etc.~as well as all proof
problems within a hierarchically structured theory database. This kind of classification knowledge
can fruitfully be employed by our agent mechanism to activate appropriate agents and especially to
deactivate non-appropriate ones.  Even if a given proof problem can not be associated with a very
restrictive class (e.g.~propositional logic) from the start, some of the subproblems subsequently
generated during the proof probably can. This can be nicely illustrated with our example: The
original proof problem belonging to higher-order logic gets transformed by the backward application
of $\Rightarrow_I$, $=_{\mbox{Subst}}$ and $\equiv$2$=$ into a very simple propositional logic
problem (cf.~line $L_4$). In this situation agents associated with a command from first- or
higher-order logic (like ${=}Subst$, $\forall E$, or {\sc Leo}\footnote{A higher-order
  theorem prover integrated to {\omega}.}) should be disabled.

Therefore we add a classification agent to our suggestion mechanism whose only task is to
investigate each new subgoal in order to classify it wrt.~to the known theories or logics. As soon
as this agent is able to associate the current goal with a known class or theory it places an
appropriate entry on the command blackboard (cf.~"HO" entry in Fig.~\ref{fig:suggestions}). This
entry is then broadcasted to the lower layer suggestion blackboards by the command agents where it
becomes available to all argument agents. Each argument agent can now compare its own classification
knowledge with the particular entry on the suggestion blackboard and decide whether it should perform
further computations within the current system state or not.

The motivation for designing the subgoal classifying component as an agent itself is clear: It can
be very costly to examine whether a given subgoal belongs to a specific theory or logic. Therefore
this task should be performed concurrently by the suggestion mechanism and not within each
initialization phase of the blackboard mechanism.  Whereas our current architecture provides
one single classification agent only, the single algorithms and tests employed by this component can
generally be further distributed by using a whole society of classification agents.


\section{Conclusion}
\label{sec:conc}

In this paper we reported on the extension of the concurrent command suggestion
mechanism~\cite{BeSo98b} to a resource adaptive approach. The resources that influence the
performance of our system are: (i) The available computation time and memory space. (ii)
Classification knowledge on the single agents and the agent societies.  (iii) Criteria and
algorithms available to the classification agent.

Our approach can be considered as an instance of a boundedly rational
system~\cite{Zilberstein95,Simon82}. The work is also related to~\cite{Gerber:Resource:98} which
presents an abstract resource concept for multi-layered agent architectures.  \cite{Jung:RoboCup:98}
describes a successful application of this framework within the Robocup simulation.  Consequently
some future work should include a closer comparison of our mechanism with this work.

The presented extensions are currently implemented and analyzed in \OMEGA\@. This might yield
further possible refinements of the resource concepts to improve the performance of the
mechanism. Another question in this context is, whether learning techniques can support our resource
adjustments on the top layer, as it seems to be reasonable that there even exist appropriate resource
patterns for the argument agents in dependence of the focused subproblem.

\paragraph{Acknowledgements} We thank Serge Autexier and Christoph Jung for stimulating discussions.

\bibliographystyle{plain}

\end{document}